\documentclass[%
 aip,
 jmp,%
 amsmath,amssymb,
preprint,%
]{revtex4-1}

\usepackage{graphicx}
\usepackage{dcolumn}
\usepackage{bm}

\begin{document}


\title[Unobservable Potentials to Explain Single Photon and Electron Interference]{Unobservable Potentials to Explain Single Photon and Electron Interference}

\author{Masahito Morimoto}
 \altaffiliation{Telecommunications \& Energy Laboratories of Furukawa Electric Co., Ltd. \\ 6, Yawata-Kaigandori, Ichihara, Chiba, 290-8555, Japan}
\email{masahito.s.morimoto@furukawaelectric.com}


\date{\today}

\begin{abstract}
We show single photon and electron interferences can be calculated without quantum-superposition states by using tensor form (covariant quantization).
From the analysis results, the scalar potential which correspond to an indefinite metric vector forms an oscillatory field and causes the interferences.
The results clarify the concept of quantum-superposition states is not required for the description of the interference, which leads to an improved understanding of the uncertainty principle and resolution of paradox of reduction of the wave packet, elimination of infinite zero-point energy and derivation of spontaneous symmetry breaking. The results conclude Quantum theory is a kind of deterministic physics without ''probabilistic interpretation''.
%
\end{abstract}

\maketitle
\section{Introduction}
\label{intro}
Basic concept of the quantum theory is the quantum-superposition states. Arbitrary states of a system can be described by pure states which are superposition of eigenstates of the system. Calculation results by the concept agree well with experiment. Without the concept, single photon or electron interference could not be explained. In addition to the interference, entangle states could not be explained, either.

However the concept leads to the paradox of the reduction of the wave packet typified by "Schr\"odinger's cat" and "Einstein, Podolsky and Rosen (EPR)" \cite{Schr-cat,EPR}. 

In order to interpret the quantum theory without paradoxes, de Broglie and Bohm had proposed so called "hidden variables" theory \cite{Bohm1,Bohm2}. Although, "hidden variables" has been negated \cite{Bell}, the theory has been extended to consistent with relativity and ontology \cite{Bohmian1,Bohmian2,Bohmian3,Bohmian4,Hiley}. However the extension has not been completed so far.

Although there were a lot of arguments about the paradoxes, recent paper related to the quantum interferences convince us of the validity of the concept.
For example, quantum mechanical superpositions by some experiments have been reviewed \cite{Arndt}.
The atom interference by using Bose-Einstein condensates (BECs) has been reported experimentally and theoretically \cite{PhysRevA.84.023619,PhysRevA.87.013632}.  
The coherence length of an electron or electron-electron interference by using the Aharonov-Bohm oscillations in an electronic MZI has been discussed theoretically \cite{PhysRevB.84.081303,PhysRevB.87.195433}. 
A plasmonic modulator utilizing an interference of coherent electron waves through the Aharonov-Bohm effect has been studied by the author \cite{6630074}.
The entangle states have been widely discussed experimentally and theoretically \cite{aspect1981experimental,aspect1982experimental,aspect1982experimental-2,PhysRevA.65.033818,PhysRevLett.100.220402,PhysRevB.79.245108}.
The photon interference by using nested MZIs and vibrate mirrors has been measured and analyzed \cite{PhysRevLett.111.240402,PhysRevA.89.033825}.
The double-slit electron diffraction has been experimentally demonstrated \cite{1367-2630-15-3-033018}.
According to our analysis, BECs, condensate and bosonization systems correspond to mixed states with or without coherence rather than pure states, and no paper has been able to solve the paradoxes. 

In this paper, we offer a new insight of the single photon and electron interference that can solve the paradoxes. According to the new insight, the description of nature does not require the concept of quantum-superposition and pure states whose probabilities are fundamental sense. Only the concept of mixed states whose probabilities are statistical sense will be justified in nature. The new insight gives us novel and important results, i,e., improved understanding of the uncertainty principle non-related to measurements, elimination of infinite zero-point energy without artificial subtraction, derivation of spontaneous symmetry breaking without complexity and knowledge that Quantum theory is a kind of deterministic physics without ''probabilistic interpretation''.

In addition, new insight can conclude that the concept of entangle state is not required though there have been reported the validity of the concept of entangle states \cite{aspect1981experimental,aspect1982experimental,aspect1982experimental-2,PhysRevA.65.033818,PhysRevLett.100.220402,PhysRevB.79.245108}. We will discuss the entangle state by using the new insight in other letter \cite{Morimoto-Entangle}.

In section \ref{sec:CE}, we show easy example of Gaussian photon beam to explain that single photon can be described by substantial (localized) photon and unobservable potentials (scalar potentials).
In section \ref{sec:PE}, we also show easy explanation that we should recognize the existence of the potentials in two-slit electron interference experiment. 
In section \ref{sec:ISP}, we show the calculation of the interferences by using tensor form which does not require quantum-superposition states. In addition to the form, we show an alternate formalism (however it's just a provisional treatment) convenient for the calculations.

In section \ref{sec:ISE}, we also show the calculation of the single electron interference in the same manner.
In section \ref{sec:DC}, we discuss the paradoxes related to quantum-superposition states, zero-point energy, spontaneous symmetry breaking and general treatment of single particle interferences. In section \ref{sec:CC}, we summarize the findings of this work.

Aharonov and Bohm had pointed out the unobservable potentials can effect the electron wave interferences and the effect had been experimentally identified by Tonomura et. al \cite{ABeffe,Tonomura1,Tonomura2}. 
The findings has pointed out the unobservable potentials (include scalar potentials) can generate not only Aharonov-Bohm effect but also single photon, electron or an arbitrary particle field interferences and fluctuation of the universe as will be described later in this paper.

These findings are obtained from the reformulation of traditional treatment of quantum theory. The mathematical tools involved in this paper such as routine state vectors, operators, inner products and so on are identical to those used in traditional quantum theory. The difference from the traditional treatment is the introduction of indefinite metric as physical substance that contradicts "probabilistic interpretation". In this reformulation, the inner product of the states which has been recognized as so called "probability amplitudes" is unrelated to the probability but related to an amplitude of interferences. Hence the "interference amplitudes" is preferable to "probability amplitudes".

The discussions and findings described in this paper are very simple from the mathematical viewpoint but rigorous and the result is an inevitable conclusion by the rigorous derivation.

\section{Classical Electromagnetic field of MZI - potentials and photon}\label{sec:CE}

\begin{figure}
\includegraphics[width=86mm]{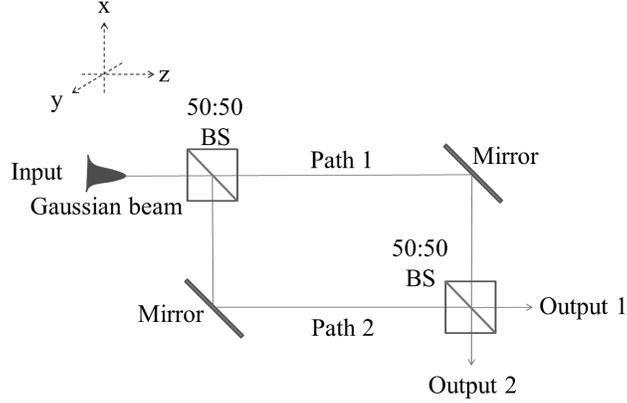}
\centering 
\caption{\label{fig:one-ph} Schematic view of MZI. BS:Beam Splitter.}
\end{figure}

Figure \ref{fig:one-ph} shows schematic view of the Mach-Zehnder Interferometer (MZI) and coordinate
system.

First we examine the input beam. Assume that an x-polarized optical beam propagates in z-direction with angular frequency $ \omega$ and propagation constant $ \beta $, the electric field $ {\bf E} $ of the optical beam is well localized in the free space, e.g., the cross section profile of the electric field is expressed as Gaussian distribution.

Then, the electric field of the optical beam in the input can be expressed as follows.
\begin{eqnarray}
{\bf E} &=& {\bf e}_{x} \cdot C_{E} \cdot \exp \left( -\frac{x^{2}+y^{2}}{w_{0}^{2}} \right) \cdot \cos \left( \omega t-\beta z \right)
\label{eq:E}
\end{eqnarray} 
where, $ {\bf e}_{x} $ is unit vector parallel to the x-axis. $ C_{E} $ is an arbitrary constant which is proportional to the magnitude of the electric field. $ w_{0} $ is the radius of the optical beam.
$ {\bf E} $ and $ {\bf B} $ are expressed by vector and scalar potentials as follows.
\begin{eqnarray}
{\bf E} &=& -\frac{\partial}{\partial t} {\bf A} - \nabla \phi \nonumber \\
{\bf B} &=&  \nabla \times {\bf A} 
\label{eq:E_potential}
\end{eqnarray} 
From (\ref{eq:E}) and (\ref{eq:E_potential}), $ {\bf A} $ is expressed by introducing a vector function $ {\bf C} $ as follows.
\begin{eqnarray}
{\bf A} & = & - \frac{1}{\omega} {\bf e}_{x} \cdot C_{E} \cdot \exp \left( -\frac{x^{2}+y^{2}}{w_{0}^{2}} \right) \cdot \sin \left( \omega t-\beta z \right)  + {\bf C} \! \nonumber \\
\frac{\partial}{\partial t}{\bf C} & = & -\nabla \phi 
\label{eq:AandC}
\end{eqnarray} 
By taking $ {\bf C} $ as an irrotational vector function $ \nabla \times {\bf C}=0 $ in order for $ {\bf B} $ to localize in the space, for example, $ {\bf C} $ and $ \phi $ can be expressed by introducing an arbitrary scalar function $ \lambda $ as $ {\bf C} = \nabla \lambda $ and $ \nabla \left( \frac{\partial}{\partial t} \lambda + \phi \right) = 0 $ respectively.

Then {\bf B} is expressed as follows
\begin{eqnarray}
{\bf B} &=& \nabla \times  {\bf A} \nonumber \\
&=& \frac{\beta}{\omega} {\bf e}_{y} \cdot C_{E} \cdot \exp \left( -\frac{x^{2}+y^{2}}{w_{0}^{2}} \right) \cdot \cos \left( \omega t-\beta z \right) \nonumber \\
& & -\frac{2y}{\omega \! \cdot \! w_{0}^{2}}  {\bf e}_{z} \! \cdot \! C_{E} \! \cdot \! \exp \! \left( \! -\frac{x^{2}+y^{2}}{w_{0}^{2}} \! \right) \! \cdot \! \sin \left( \omega t-\beta z \right)
\label{eq:B}
\end{eqnarray} 

Therefore, $ {\bf E} $ and $ {\bf B} $ are localized in the free space in the input. In contrast, the vector and scaler potentials, which can not be observed alone, are not necessarily localized. Especially the scalar potentials have no effect on the $ {\bf E} $ and $ {\bf B} $.

Note that, the Gaussian beam radius will be spatially expanded due to the free space propagation. However, the radius of the propagated beam $ w \left( z \right) $ will be approximately 10.5mm when the beam with the initial radius $ w_{0} = 10 $mm propagates $ z=100 $m in free space. This value can be calculated by $ w \left( z \right) = w_{0} \sqrt{1+\left( \frac{\lambda z}{\pi w_{0}^{2}} \right)^{2}} $ when the wavelength $ \lambda = 1 \mu $m is applied. Then the spatially expansion of the beam will be negligible small when the paths of the MZI are less than several tens meters.

The above localized form (\ref{eq:E}) is one example, other forms which satisfy the following Maxwell equations can be employed.
\begin{eqnarray}
\left( \Delta -\frac{1}{c^{2}} \frac{\partial ^{2}}{\partial t^{2}} \right) {\bf A}- \nabla \left( \nabla \cdot {\bf A} + \frac{1}{c^{2}} \frac{\partial \phi}{\partial t} \right) =- \mu_{0} {\bf i} \nonumber \\
\left( \Delta -\frac{1}{c^{2}} \frac{\partial ^{2}}{\partial t^{2}} \right) \phi + \frac{\partial}{\partial t} \left( \nabla \cdot {\bf A} + \frac{1}{c^{2}} \frac{\partial \phi}{\partial t} \right) = - \frac{\rho}{\varepsilon_{0}}
\label{eq:Maxi}
\end{eqnarray}
where $ \mu_{0} $ is the permeability and $ \rho $ is the electric charge density.

When $ {\bf i} = 0 $ and $ \rho = 0 $, the equations (\ref{eq:Maxi}) can express the localized electromagnetic fields in free space as described in the above.

\section{Interference of single photon}\label{sec:ISP}

As described previously, there are potentials which are not necessarily localized even if photons are localized. Especially the scalar potential can populate the whole of space and the vector and scalar potentials are combined by Lorentz transformation. Then we should make no distinction between the vector and scalar potentials.  

However traditional treatment of the single photon interference by using Coulomb gauge only uses the quantized vector potentials as follows. 
In a quantum mechanical description, the photon interference is calculated by introducing the electric field operator $ \hat{E}=\frac{1}{\sqrt{2}} \hat{a}_{1} \exp \left( i \theta \right) + \frac{1}{\sqrt{2}} \hat{a}_{2} $ and the number state $ | n \rangle $ of the MZI output as follows \cite{Loudon}. Where $ \hat{a}_{\rm 1 or 2} $ is the photon annihilation operator corresponding to an optical mode passing through path 1 or 2, respectively, $ \theta $ is the phase difference corresponding to the difference in length between the two paths.
\begin{eqnarray}
\langle \hat{I} \rangle & \propto & \langle n | \hat{E}^{\dagger} \hat{E} | n \rangle \nonumber \\
& = & \frac{1}{2} \langle n | \hat{a}_{1}^{\dagger} \hat{a}_{1} | n \rangle + \frac{1}{2} \langle n | \hat{a}_{2}^{\dagger} \hat{a}_{2} | n \rangle +\cos \theta \langle n | \hat{a}_{1}^{\dagger} \hat{a}_{2} | n \rangle
\label{eq:trad-int}
\end{eqnarray}
Where $ \langle \hat{I} \rangle $ is expectation value of the field intensity which is proportional to photon number. $ \hat{a}_{\rm 1 or 2} $ and $ \hat{a}_{\rm 1 or 2}^{\dagger} $ are defined as $ \hat{a} = \frac{\hat{a}_{\rm 1} + \hat{a}_{\rm 2}}{\sqrt{2}} $ and $ \hat{a}^{\dagger} = \frac{\hat{a}_{\rm 1}^{\dagger} + \hat{a}_{\rm 2}^{\dagger}}{\sqrt{2}} $ by using the photon annihilation and creation operators $ \hat{a} $ and $ \hat{a}^{\dagger} $at the input with $ \langle n | \hat{a}_{1}^{\dagger} \hat{a}_{1} | n \rangle =  \langle n | \hat{a}_{2}^{\dagger} \hat{a}_{2} | n \rangle = \langle n | \hat{a}_{1}^{\dagger} \hat{a}_{2} | n \rangle = \frac{1}{2} n $. When photon number is one ( $ n=1 $ ), i.e., single photon, the above expectation value is calculated to be $ \langle \hat{I} \rangle \propto  \frac{1}{4} + \frac{1}{4} + \frac{1}{2} \cos \theta = \frac{1}{2}+ \frac{1}{2} \cos \theta $.

In this traditional  treatment, the photon annihilation and creation operators $ \hat{a} $ and $ \hat{a}^{\dagger} $ are obtained from quantization of the vector potentials in (\ref{eq:Maxi}) by using Coulomb gauge under assumption of $ {\bf i} = 0 $ and $ \rho = 0 $. 

In order to equate scalar potentials with vector potentials, i. e., $ \langle 1 | \hat{A}_{0}^{\dagger} \hat{A}_{0} | 1 \rangle = \langle 1 | \hat{A}_{1}^{\dagger} \hat{A}_{1} | 1 \rangle $ as will be described later, we should introduce tensor form (covariant quantization) as follows.

The electromagnetic potentials are expressed as following four-vector in Minkowski space.
\begin{equation}
A^{\mu} = (A^{0}, \ A^{1}, \ A^{2}, \ A^{3} ) = (\phi /c, \ {\bf A})
\end{equation}
The  four-current are also expressed as following four-vector.
\begin{equation}
j^{\mu} = (j^{0}, \ j^{1}, \ j^{2}, \ j^{3}) = (c \rho, \ {\bf i})
\end{equation}
When we set the axises of Minkowski space to $ x^{0} = ct $, $ x^{1} = x, \ x^{2} = y, \ x^{3}  =z $, Maxwell equations with Lorentz condition are expressed as follows.
\begin{eqnarray}
\Box A^{\mu} & = & \mu_{0} j^{\mu} \nonumber \\
\partial_{\mu} A^{\mu} & = & 0
\end{eqnarray}
In addition, the conservation of charge $ {\rm div} \ {\bf i} + \partial \rho / \partial t = 0 $ is expressed as $ \partial_{\mu} j^{\mu} = 0 $.
Where $ \partial_{\mu} = ( 1/c\partial t , \ 1/\partial x , \ 1/\partial y , \ 1/\partial z ) = ( 1/\partial x^{0} , \ 1/\partial x^{1} , \ 1/\partial x^{2} , \ 1/\partial x^{3} ) $ and  $ \Box $ stands for the d'alembertian: $ \Box \equiv \partial_{\mu} \partial^{\mu} \equiv \partial^2 / c^2 \partial t^2 - \Delta $.

The transformation between covariance and contravariance vector can be calculated by using the simplest form of Minkowski metric tensor $ \texttt{g}_{\mu \nu} $ as follows.
\begin{eqnarray}
 \texttt{g}_{\mu \nu} = \texttt{g}^{\mu \nu} & = & \left[ 
 \begin{array}{cccc}
 1 & 0 & 0 & 0 \\
0 & -1 & 0 & 0 \\
0 & 0 & -1 & 0 \\
0 & 0 & 0 & -1\\
 \end{array} 
\right] \nonumber \\
A_{\mu} & = & \texttt{g}_{\mu \nu} A^{\nu} \nonumber \\
A^{\mu} & = & \texttt{g}^{\mu \nu} A_{\nu}
\end{eqnarray} 
The following quadratic form of four-vectors is invariant under a Lorentz transformation.
\begin{equation}
(x^{0} )^{2} - ( x^{1} )^{2} - (x^{2} )^{2} - (x^{3} )^{2} 
\end{equation}
The above quadratic form applied a minus sign expresses the wave front equation and can be described by using the metric tensor.

\begin{equation}
- \texttt{g}_{\mu \nu} x^{\mu} x^{\nu} = - x^{\mu} x_{\mu} = x^{2} + y^{2} + z^{2} - c^{2} t^{2} = 0
\end{equation}
This quadratic form which includes minus sign is also introduced to inner product of arbitrarily vectors and commutation relations in Minkowski space.

The four-vector potential satisfied Maxwell equations with vanishing the four-vector current can be expressed as following Fourier transform in terms of plane wave solutions \cite{QFT}.
\begin{equation}
A_{\mu} (x) = \int d \tilde{k} \sum^{3}_{\lambda = 0} [a^{(\lambda)} (k) \epsilon^{(\lambda)}_{\mu} (k) e^{- i k \cdot x } + a^{(\lambda) \dagger} (k) \epsilon^{(\lambda) *}_{\mu} (k) e^{ i k \cdot x} ]
\end{equation}
\begin{equation}
\tilde{k} = \frac{d^{3} k}{2 k_{0} (2 \pi )^{3}} \ \ \ k_{0} = | {\bf k} | 
\end{equation}
where the unit vector of time-axis direction $ n $ and polarization vectors $ \epsilon^{(\lambda )}_{\mu} (k) $ are introduced as $ n^{2} = 1, \ n^{0}>0 $ and $ \epsilon^{(0)} = n $, $ \epsilon^{(1)} $ and $ \epsilon^{(2)} $ are in the plane orthogonal to $k$ and $n$
\begin{equation}
\epsilon^{(\lambda)} (k) \cdot \epsilon^{(\lambda ')} (k) = - \delta_{\lambda , \lambda '} \ \ \ \  \lambda \ , \  \lambda ' = 1, \ 2
\end{equation}
$ \epsilon^{(3)} $ is in the plane $ (k, \ n) $ orthogonal to $n$ and normalized
\begin{equation}
\epsilon^{(3)} (k) \cdot n = 0 \ , \ [ \epsilon^{(3)} (k) ]^2 = -1
\end{equation}

Then $ \epsilon^{(0)} $ can be recognized as a polarization vector of scalar waves, $ \epsilon^{(1)} $ and $ \epsilon^{(2)} $ of transversal waves and $ \epsilon^{(3)} $ of a longitudinal wave. Then we take these vectors as following the easiest forms.
\begin{equation}
\epsilon^{(0)} = \left(
\begin{array}{l}
1 \\
0 \\
0 \\
0
\end{array}
\right)
\ \ \ 
\epsilon^{(1)} = \left(
\begin{array}{l}
0 \\
1 \\
0 \\
0
\end{array}
\right)
\ \ \ 
\epsilon^{(2)} = \left(
\begin{array}{l}
0 \\
0 \\
1 \\
0
\end{array}
\right)
\ \ \ 
\epsilon^{(3)} = \left(
\begin{array}{l}
0 \\
0 \\
0 \\
1
\end{array}
\right)
\end{equation}
For simplicity, assume that photons are x-polarized transversal waves with the scalar wave and we neglect the longitudinal wave which is considered to be unphysical presence, i. e., $\ A_{2} = 0, \ A_{3} = 0 $.
\begin{equation}
A_{\mu} = (A_{0}, \ A_{1}, \ 0, \ 0)
\label{eq:A_mu}
\end{equation}

The potentials will be divided when there are two paths divided by the MZI interferometer. Here we consider the state that a photon expressed as x-polarized transversal waves passes through path 1 and unobservable potentials, i. e., $ A_{0} (x) $, is divided into both path 1 and 2 with a phase difference between the two paths. In this state, we can express the four-vector potentials along the MZI path 1 ( $ \equiv A_{\mu : ({\rm path 1})} $ ) and path 2 ( $ \equiv A_{\mu : ({\rm path 2})} $ ) as follows.
\begin{eqnarray}
 A_{\mu : ({\rm path 1})}  & = & (\frac{1}{2} e^{i \theta / 2 } A_{0} , \ A_{1} , \ 0 , \ 0 ) \nonumber \\
 A_{\mu : ({\rm path 2})}  & = & (\frac{1}{2} e^{- i \theta / 2 } A_{0} , \ 0, \ 0, \ 0 ) 
\label{eq:appMZIA}
\end{eqnarray}

When the Fourier coefficients of the four-vector potentials are replaced by operators as $ \hat{A}_{\mu} \equiv \sum^{3}_{\lambda = 0} \hat{a}^{(\lambda)} (k) \epsilon^{(\lambda)}_{\mu} (k) $, the commutation relations are obtained as follows.
\begin{equation}
[ \hat{A}_{\mu} (k), \ \hat{A}_{\nu}^{\dagger} (k') ] = - \texttt{g}_{\mu \nu} \delta (k-k')
\label{eq:comm}
\end{equation}
The time-axis component (corresponds to $ \mu , \nu = 0 $ scalar wave, i. e., scalar potential because $ \epsilon^{(0)}_{\mu} (k) = 0 \ (\mu \neq 0)  $) has the opposite sign of the space axes. Because $ \langle 0 | \hat{A}_{0} (k)  \hat{A}_{0}^{\dagger} (k') | 0 \rangle = - \delta (k-k') $ then
\begin{equation}
\langle 1 | 1 \rangle = - \langle 0 | 0 \rangle \int d \tilde{k} | f(k) |^{2}
\end{equation}
where $ | 1 \rangle = \int d \tilde{k} f(k) \hat{A}_{0}^{\dagger} (k) | 0 \rangle $. Therefore the time-axis component is the root cause of indefinite metric. In order to utilize the indefinite metric as followings, Coulomb gauge that removes the scalar potentials should not be used.

Let define the photon annihilation operators $ \hat{A}_{\mu : ({\rm path 1})}  $ and $ \hat{A}_{\mu : ({\rm path 2})}  $ corresponding to the optical modes passing through the MZI path 1 and 2 respectively. The products of these operators must introduce the same formalism.
\begin{equation}
\hat{A}^{\dagger} \hat{A} = - \texttt{g}_{\mu \nu} \hat{A}^{\mu \dagger} \hat{A}^{\nu} = - \texttt{g}^{\mu \nu} \hat{A}_{\mu}^{\dagger} \hat{A}_{\nu}
\end{equation}

Because the photon annihilation operator at the MZI output is $ \hat{A}_{\mu : ({\rm path 1})} + \hat{A}_{\mu : ({\rm path 2})} $, then we can obtain the photon number operator at the MZI output as follows.
\begin{eqnarray}
& & - \texttt{g}^{\mu \nu} \{\hat{A}_{\mu : ({\rm path 1})} + \hat{A}_{\mu : ({\rm path 2})}  \}^{\dagger} \{ \hat{A}_{\nu : ({\rm path 1})} + \hat{A}_{\nu : ({\rm path 2})}  \} \nonumber \\
& = & - \frac{1}{2} \hat{A}_{0}^{\dagger} \hat{A}_{0} + \hat{A}_{1}^{\dagger} \hat{A}_{1} -\frac{1}{2} \hat{A}_{0}^{\dagger} \hat{A}_{0} \cos \theta
\label{eq:appinter}
\end{eqnarray}
where the following relations are used.
\begin{eqnarray}
\begin{array}{lll}
- \texttt{g}^{\mu \nu} \hat{A}_{\mu : ({\rm path 1})} ^{\dagger} \hat{A}_{\nu : ({\rm path 1})} & = & -\frac{1}{4} \hat{A}_{0}^{\dagger} \hat{A}_{0} + \hat{A}_{1}^{\dagger} \hat{A}_{1} \nonumber \\
- \texttt{g}^{\mu \nu} \hat{A}_{\mu : ({\rm path 1})} ^{\dagger} \hat{A}_{\nu : ({\rm path 2})}  & = & -\frac{1}{4} e^{- i \theta} \hat{A}_{0}^{\dagger} \hat{A}_{0} \nonumber \\
- \texttt{g}^{\mu \nu} \hat{A}_{\mu : ({\rm path 2})} ^{\dagger} \hat{A}_{\nu : ({\rm path 1})}  & = & -\frac{1}{4} e^{i \theta} \hat{A}_{0}^{\dagger} \hat{A}_{0} \nonumber \\
- \texttt{g}^{\mu \nu} \hat{A}_{\mu : ({\rm path 2})} ^{\dagger} \hat{A}_{\nu : ({\rm path 2})}  & = & -\frac{1}{4} \hat{A}_{0}^{\dagger} \hat{A}_{0}
\end{array}
\end{eqnarray}

Applying the bra and ket vectors $ \langle 1 | $ and $ | 1 \rangle $ to (\ref{eq:appinter}),  $ \langle \hat{I} \rangle \propto \frac{1}{2} - \frac{1}{2} \cos \theta $ is obtained.
Note that we identify the number operators as $ \langle 1 | \hat{A}_{0}^{\dagger} \hat{A}_{0} | 1 \rangle = \langle 1 | \hat{A}_{1}^{\dagger} \hat{A}_{1} | 1 \rangle = 1 $ because of the Lorentz invariance.

From the time-reversal invariance of the electromagnetic fields, we should also make no distinction between the input and output of the MZI. Then the photon annihilation operator at the confluence of the MZI can be expressed as same as the MZI output, i. e.,
\begin{equation}
 \hat{A}_{\mu} = \hat{A}_{\mu : ({\rm path 1})} + \hat{A}_{\mu : ({\rm path 2})} = \left( \cos \frac{\theta}{2} \hat{A}_{0} , \ \hat{A}_{1} , \ 0 , \ 0 \right) 
\label{eq:A_conf}
\end{equation}
instead of (\ref{eq:A_mu}). Although there is definitely a photon at the MZI input, the calculation result of the photon number at the MZI input of a single photon state by using (\ref{eq:A_mu}) is $ \langle 1 | (- \hat{A}_{0}^{\dagger} \hat{A}_{0} + \hat{A}_{1}^{\dagger} \hat{A}_{1} ) | 1 \rangle = 0 $. However we should not omit the scalar potentials as $ \hat{A}_{0} = 0 $.

In contrast, the photon number by using (\ref{eq:A_conf}) with $ \theta = \pm N \pi \ (N : {\rm odd \ number} ) $ is 1. Therefore we should recognize the scalar potentials at the MZI input are not zero (not empty, i. e., $ \hat{A}_{0} \neq 0 $) but annihilate each other by the opposite phase waves, i. e., $ \cos (\theta/2) = 0 $.
When there are two paths, the scalar potentials make oscillatory fields like $ f ( \theta ) \cdot \hat{A}_{0} $ where $ f ( \theta ) $ is an oscillating function of $ \theta $, which can be recognized as "hidden variables". Then the substantial photons move with the interference in the oscillatory fields. Therefore the expectation value of the field intensity at an arbitrary point in space can be calculated using (\ref{eq:A_conf}) as $ \langle \hat{I} \rangle \propto \frac{1}{2} - \frac{1}{2} \cos \theta $ which means even if the substantial photon follows an arbitrary path the photon can not be found at the point whose $ \theta = \pm N \pi \ (N : {\rm even \ number} ) $ on the path. Note that if $ \hat{A}_{1} = 0 $, i. e., there are only scalar potentials, the intensity of the oscillatory field at an arbitrary point in space negatively fluctuates like $ \langle \hat{I} \rangle \propto - \frac{1}{2} - \frac{1}{2} \cos \theta $.

The tensor form (\ref{eq:appMZIA}) can offer clear image that the substantial photon passes through one side path of the MZI and there are the unobservable potentials (scalar potentials) at both paths. As the above calculation shows the unobservable potentials act as a homodyne local oscillator which retrieves phase information from a signal (photon) through interference between the signal and local oscillator.

If we introduce following operator $ \hat{A}_{0}' $ by using the above operator $ \hat{A}_{1} $,  we can calculate the MZI interference based on Heisenberg picture without tensor form. Although the following formalism is just a provisional treatment, it is convenient for the calculations.

\begin{eqnarray}
\hat{A}_{0}' & = & \frac{1}{2} \gamma e^{i \theta/2} \hat{A}_{1} - \frac{1}{2} \gamma e^{- i \theta/2} \hat{A}_{1} \nonumber \\
\hat{A}_{0}'^{\dagger} & = & \frac{1}{2} \gamma e^{- i \theta/2} \hat{A}_{1}^{\dagger} - \frac{1}{2} \gamma e^{ i \theta/2} \hat{A}_{1}^{\dagger}
\label{eq:a2}
\end{eqnarray}
where $ \gamma^{2} = - 1 $ ( i. e., $ \gamma $ corresponds to the square root of the determinant of Minkowski metric tensor $ \sqrt{ | \texttt{g}_{\mu \nu} | } \equiv \sqrt{\texttt{g}} \equiv \sqrt{-1} = \gamma $.) $ \hat{A}_{0}' $ correspond to $ \hat{A}_{0} $ (scalar potential which express the homodyne local oscillator) in (\ref{eq:A_conf}), though the correspondence is not completely compatible with  the tensor form because of a provisional treatment, e. g., the phase is $ \pi $ shifted as described later.

Then by using this operator, the expectation value of the field intensity $ \langle \hat{I} \rangle \propto \langle 1 | (\hat{A}_{0}' + \hat{A}_{1} )^{\dagger} (\hat{A}_{0}' + \hat{A}_{1} ) | 1 \rangle $ can be calculated as follows.

\begin{eqnarray}
\hat{A}_{0}'^{\dagger} \hat{A}_{0}' & = & - \frac{1}{4} \hat{A}_{1}^{\dagger} \hat{A}_{1} - \frac{1}{4} \hat{A}_{1}^{\dagger} \hat{A}_{1} + \frac{1}{4} e^{ i \theta} \hat{A}_{1}^{\dagger} \hat{A}_{1} + \frac{1}{4} e^{- i \theta} \hat{A}_{1}^{\dagger} \hat{A}_{1}\nonumber \\
& = & - \frac{1}{2} \hat{A}_{1}^{\dagger} \hat{A}_{1} + \frac{1}{2} \hat{A}_{1}^{\dagger} \hat{A}_{1} \cos \theta \nonumber \\
\hat{A}_{1}^{\dagger} \hat{A}_{0}' & = & \frac{1}{2} \gamma e^{i \theta/2} \hat{A}_{1}^{\dagger} \hat{A}_{1} - \frac{1}{2} \gamma e^{- i \theta/2} \hat{A}_{1}^{\dagger} \hat{A}_{1} \nonumber \\
\hat{A}_{0}'^{\dagger} \hat{A}_{1} & = & \frac{1}{2} \gamma e^{- i \theta/2} \hat{A}_{1}^{\dagger} \hat{A}_{1} - \frac{1}{2} \gamma e^{ i \theta/2} \hat{A}_{1}^{\dagger} \hat{A}_{1}
\label{eq:com}
\end{eqnarray}
Finally the following result is obtained.
\begin{eqnarray}
\langle 1 | \hat{A}_{1}^{\dagger} \hat{A}_{1} | 1 \rangle & = & 1 \nonumber \\
\langle 1 | \hat{A}_{0}'^{\dagger} \hat{A}_{0}' | 1 \rangle & = & - \frac{1}{2} + \frac{1}{2} \cos \theta \nonumber \\
\langle 1 | \hat{A}_{1}^{\dagger} \hat{A}_{0}' | 1 \rangle & = & \frac{1}{2} \gamma e^{i \theta/2}  - \frac{1}{2} \gamma e^{- i \theta/2} \nonumber \\
\langle 1 | \hat{A}_{0}'^{\dagger} \hat{A}_{1} | 1 \rangle & = &  \frac{1}{2} \gamma e^{- i \theta/2} - \frac{1}{2} \gamma e^{ i \theta/2} \nonumber
\end{eqnarray}
\begin{equation}
\langle 1 | \hat{A}_{1}^{\dagger} \hat{A}_{1} | 1 \rangle + \langle 1 | \hat{A}_{0}'^{\dagger} \hat{A}_{0}' | 1 \rangle + \langle 1 | \hat{A}_{1}^{\dagger} \hat{A}_{0}' | 1 \rangle + \langle 1 | \hat{A}_{0}'^{\dagger} \hat{A}_{1} | 1 \rangle = \frac{1}{2} + \frac{1}{2} \cos \theta
\label{eq:inter}
\end{equation}
Note that when we use this provisional treatment instead of the tensor form, the phase is $ \pi $ shifted.

This provisional treatment will correspond to using the following tensor form instead of (\ref{eq:appMZIA}).

\begin{eqnarray}
\hat{A}_{\mu : ({\rm path 1})} & \equiv & (0, \ \hat{A}_{1}, \ 0, \ 0 ) \nonumber \\
\hat{A}_{\mu : ({\rm path 2})}  & \equiv & ( \frac{1}{2} i e^{i \theta /2 } \hat{A}_{0} - \frac{1}{2} i e^{-i \theta /2} \hat{A}_{0}, \ 0, \ 0, \ 0)
\label{eq:appMZIA-defmet}
\end{eqnarray}

The above calculation is based on Heisenberg picture. We can obtain the same interference based on Schr\"odinger picture. In Schr\"odinger picture, the expectation value of the field intensity can be calculated by using the output 1 (or 2: $ \frac{\pi}{2} $ phase difference) state $ | 1 \rangle_{S} + | \zeta \rangle $ and a photon annihilation operator of Schr\"odinger picture $ \hat{A}_{S} $ which is proportional to the electric field operator $ \hat{E} \propto \hat{A}_{S} $ at the output 1 (or 2). 
Where $ | 1 \rangle_{S} $ and $ | \zeta \rangle $ represent the states of a photon passes through path 1 and unobservable potentials (scalar potentials) passes through (exists at) path 2 respectively.
Because nothing is observed in path 2, we should recognize $ \langle \zeta | \zeta \rangle = 0 $. More precise definition is as follows. The operators $ \hat{A}_{1} $, $ \hat{A}_{S} $ and states $ | 1 \rangle $, $ | 1 \rangle_{S} $ can be translated by using the Hamiltonian $ \hat{\mathcal{H}} $ as $ \hat{A}_{1} = e^{i \hat{\mathcal{H}} t / \hbar} \hat{A}_{S}  e^{- i \hat{\mathcal{H}} t / \hbar} $ and $ | 1 \rangle_{S} = e^{ - i \hat{\mathcal{H}} t / \hbar} | 1 \rangle $ respectively. Then $ \hat{A}_{0}' | 1 \rangle $ can be expressed by using (\ref{eq:a2}) as follows.
\begin{eqnarray}
\hat{A}_{0}' | 1 \rangle & = & e^{i \hat{\mathcal{H}} t / \hbar} \hat{A}_{S} \left( \frac{1}{2} \gamma e^{i \theta/2} e^{- i \hat{\mathcal{H}} t / \hbar} - \frac{1}{2} \gamma e^{- i \theta/2} e^{- i \hat{\mathcal{H}} t / \hbar} \right) | 1 \rangle \nonumber \\
& = &  e^{i \hat{\mathcal{H}} t / \hbar} \hat{A}_{S} \left( \frac{1}{2} \gamma e^{i \theta/2} - \frac{1}{2} \gamma e^{- i \theta/2} \right) | 1 \rangle_{S} 
\label{eq:zetasta}
\end{eqnarray}
Here we define 
\begin{equation}
| \zeta \rangle \equiv \left( \frac{1}{2} \gamma e^{i \theta/2} - \frac{1}{2} \gamma e^{- i \theta/2} \right) | 1 \rangle_{S} 
\label{eq:zeta}
\end{equation}
Hence $ \langle 1 | \hat{A}_{0}'^{\dagger} \hat{A}_{0}' | 1 \rangle = \langle \zeta |  \hat{A}^{\dagger}_{S} \hat{A}_{S} | \zeta \rangle $. When $ \theta = 0 $, $ | \zeta \rangle = 0 $, i. e., $ \langle \zeta | \zeta \rangle = 0 $.
In this picture, the expectation value can be expressed as follows.
\begin{eqnarray}
\langle \hat{I} \rangle & \propto & \left( \langle 1 |_{S} + \langle \zeta | \right) \hat{A}^{\dagger}_{S} \hat{A}_{S} \left( | 1 \rangle_{S} + | \zeta \rangle \right) \nonumber \\
& = & 1 + \langle \zeta | \hat{A}^{\dagger}_{S} \hat{A}_{S} | \zeta \rangle + \langle 1 | \zeta \rangle_{S} + \langle \zeta | 1 \rangle_{S} \nonumber \\
& = & 1 - \frac{1}{2} +\frac{1}{2} \cos \theta = \frac{1}{2} +\frac{1}{2} \cos \theta
\label{eq:new-int-sch}
\end{eqnarray}

In the above mathematical formula for the interference by Schr\"odinger picture, there is no mathematical solution in usual Hilbert space. Therefore the unobservable potentials (scalar potentials), which can not be observed alone, must be regarded as a vector in indefinite metric Hilbert space as can be seen from (\ref{eq:zeta}). Although the explicit expression such as (\ref{eq:zeta}) has not been reported, the same kind of unobservable vector has been introduced as "ghost" in quantum field theory \cite{Dirac,Pauli,Gupta,Lee} .
We also call $ | \zeta \rangle $ "ghost" in this paper though this "ghost" has a different definition. The traditional "ghost" was introduced mathematically as an auxiliary field for consistent with relativistic covariance of the theory and had no effect on physical phenomena. However, the above "ghost" is a physical substance (corresponds to the scalar potentials in (\ref{eq:appMZIA})) which causes the interferences, in other words, is essential for the interferences instead of the mathematical auxiliary field. 

From the equation (\ref{eq:new-int-sch}), the unobservable potentials pass through path 2 produce the single photon interference as if the photon passes through the both paths in cooperation with a photon field passes through path 1.

This discussion can be generalized for arbitrary geometries include for the above 2-paths MZI. The arbitrary geometries can be modeled by using split coefficients $ r_{j} $ and phases $ \theta_{j} $ of the scalar potential. When there are multiple path (M paths), the scalar potentials can be divided as follows.

\begin{equation}
\sum_{j=1}^{M} r_{j} e^{i \theta_{j}} A_{0}
\label{eq:divid}
\end{equation}
where $ \sum_{j=1}^{M} r_{j} = 1 $. The above MZI case corresponds to $ M = 2 $, $ r_{1} = r_{2} = 1/2 $ and $ \theta_{1} = - \theta_{2}= \theta/2 $. Then we can predict the expectation value for arbitrary geometries can be calculated by using the photon annihilation operator as follows.
\begin{equation}
 \hat{A}_{\mu} = \left( \sum_{j=1}^{M} r_{j} e^{i \theta_{j}} \hat{A}_{0} , \ \hat{A}_{1} , \ 0 , \ 0 \right) 
\label{eq:divid-inf}
\end{equation}
Then
\begin{eqnarray}
\langle \hat{I} \rangle &  \propto & \langle 1 | - \texttt{g}^{\mu \nu} \hat{A}_{\mu}^{\dagger} \hat{A}_{\nu} | 1 \rangle \nonumber \\
& = & - \left\{ \sum_{j = 1 , \ k =1 }^{M} r_{j} r_{k} e^{i ( \theta_{j} - \theta_{k})} \right\} \langle 1 |\hat{A}_{0}^{\dagger} \hat{A}_{0} | 1 \rangle + \langle 1 | \hat{A}_{1}^{\dagger} \hat{A}_{1} | 1 \rangle \nonumber \\
& = &  - \left\{ ( r_{1}^{2} + r_{2}^{2} + \cdot + r_{M}^{2} ) + \sum_{j \neq k}^{M} r_{j} r_{k} e^{i ( \theta_{j} - \theta_{k})} \right\} + 1
\label{eq:arb-number}
\end{eqnarray}
Because $ 0 \leqq r_{j} \leqq 1$, then $ 0 \leqq \langle 1 | - \texttt{g}^{\mu \nu} \hat{A}_{\mu}^{\dagger} \hat{A}_{\nu} | 1 \rangle \leqq 1 $.

When $ M \rightarrow \infty $, the multi path can be recognized as a continuum space. Because $ \sum_{j=1}^{\infty} r_{j} e^{i \theta_{j}} $ that creates the oscillatory field converges with 0 when the phases are completely random, the real physical space (we refer to this space as ''real vacuum''. In contrast we refer to the space with $ M=1 $ in (\ref{eq:divid}) as ''ideal vacuum''.) can be recognized as the continuum with completely random phases. In this case $ \langle \hat{I} \rangle \propto 1 $. When a particular geometry is formed in the space, the expectation value fluctuates by the oscillatory field.

The expression (\ref{eq:divid}) is similar to a normalized quantum-superposition state if we identify $ r_{j} $ and $ e^{i \theta_{j}} A_{0} $ as a normalization coefficient and eigenfunction (eigenstate) respectively, though  $ \sum_{j=1}^{M} r_{j} = 1 $ instead of commonly used $ \sum_{j=1}^{M} | r_{j} |^{2} = 1 $. Then we should recognize what forms the quantum-superposition-like (not completely the same expression) is not a substantial photon but the unobservable scalar potential. 

In case of the provisional treatment, we can apply following operators and state to an arbitrary geometry instead of (\ref{eq:a2}) and (\ref{eq:zeta}).

\begin{eqnarray}
\hat{A}_{0}' & = & \gamma \sum_{j=1}^{\infty} r_{j} e^{i \theta_{j}} \hat{A}_{1} - \gamma \sum_{j=1}^{\infty} r_{j} e^{- i \theta_{j}} \hat{A}_{1} \nonumber \\
\hat{A}_{0}'^{\dagger} & = & \gamma \sum_{j=1}^{\infty} r_{j} e^{- i \theta_{j}} \hat{A}_{1}^{\dagger} - \gamma \sum_{j=1}^{\infty} r_{j} e^{i \theta_{j}} \hat{A}_{1}^{\dagger}
\label{eq:a2-arb}
\end{eqnarray}

\begin{equation}
| \zeta \rangle \equiv \left( \gamma \sum_{j=1}^{\infty} r_{j} e^{i \theta_{j}} - \gamma \sum_{j=1}^{\infty} r_{j} e^{- i \theta_{j}} \right) | 1 \rangle_{S} 
\label{eq:zeta-arb}
\end{equation}

Note that the superposition principle may be used as a nice mathematical tool to simplify analyses in mixed states. However we should investigate whether engineering applications based on reduction of wave packet are feasible or not, because even single photon interference can be described without quantum superposition as described above.

\section{Potentials and electron}\label{sec:PE}

\begin{figure}
\includegraphics[width=86mm]{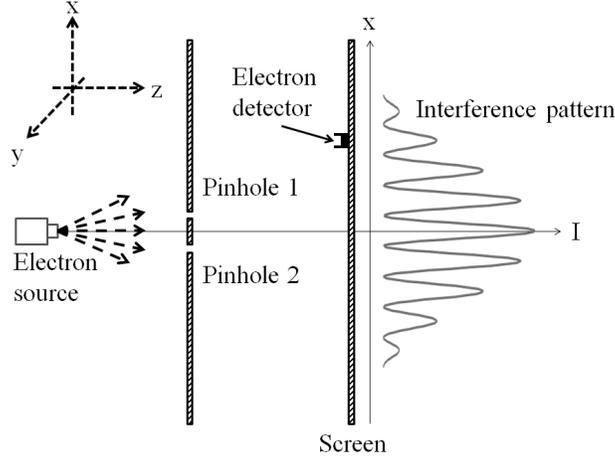}
\centering 
\caption{\label{fig:one} Schematic view of a typical setup for the 2-slits (2-pinholes) single electron interference experiment.}
\end{figure}

In this section, we first consider two pinholes electron wave interference in classical manner.
Figure \ref{fig:one} shows schematic view of a typical setup for the 2-slits (2-pinholes) single electron interference experiment \cite{1367-2630-15-3-033018,Feynman-1}.

The propagating electron can be identified as an electron beam whose space current density is $ j = N q v $, where $ N $ is the number of electron per unit volume, $q$ is the electron charge and $v$ is the electron velocity. When  the radius of the electron beam is $ w_{0} $, the current $ I $ can be expressed as $ I = \pi w_{0}^{2} j $. According to Biot-Savart Law, the propagation generates magnetic fields and potentials around the propagation path.

Assume that the electron propagates parallel to z-axis at a constant velocity.
Then, the vector potentials around the propagation path are expressed as \cite{Feynman-1,Stratton}
\begin{eqnarray}
A_{x} & = & A_{y} = 0 \nonumber \\
A_{z} & = & \frac{I}{2 \pi \varepsilon_{0} c^{2}} \ln \frac{1}{r} 
\label{eq:Vectorpotential}
\end{eqnarray} 
where $ r= \sqrt{x^{2}+y^{2}} $, $ \varepsilon_{0} $ is the permittivity and $ c $ is the speed of light.

Therefore the vector potential clearly passes through not only the pinhole the electron passes through but also the opposite pinhole.

Even if we apart from this easy consideration, the electron motion definitely generates potentials. Therefore, when we consider the electron motion, we must take the potentials. 

In next section, we consider the two pinholes interference in quantum mechanical manner with consideration for the potentials.

\section{Interference of single electron}\label{sec:ISE}

In a traditional quantum mechanical description, the 2-slits (pinholes) single electron interference is typically explained by the probability (density) of finding the electron on the screen \cite{Feynman-1}.
\begin{equation}
P_{12} = | \phi_{1} + \phi_{2} |^{2}
\label{eq:trad-int-ele}
\end{equation}
Where $ \phi_{1} = \langle x | 1 \rangle \langle 1 | s \rangle $ and  $ \phi_{2} = \langle x | 2 \rangle \langle 2 | s \rangle $, which are composed of probability amplitudes

$ \langle 1_{\rm or} 2 | s \rangle $: "$ \langle $electron arrives at pinhole 1 or 2$ | $electron leaves $ s $ (electron source)$ \rangle $" and

$ \langle x | 1_{\rm or} 2\rangle $: "$ \langle $electron arrives at screen $x | $electron leaves pinhole 1 or 2$ \rangle $".

When either pinhole 1 or 2 is closed, the each and total probabilities are calculated to be $ P_{1}=| \phi_{1}|^{2} $, $ P_{2}=| \phi_{2}|^{2} $ and $ P=P_{1}+P_{2} \neq P_{12} $.
Therefore we can not help but admit the electron passes through both pinholes at the same time despite an electron can not be split off, which requires the introduction of the quantum-superposition states .

However we can examine the states of the localized electron propagation and unobservable potentials instead of the quantum-superposition state as mentioned above.

In such a case, the electron wave functions should be expressed as follows.
\begin{eqnarray}
\psi_{1}' &=& \psi_{1} \cdot  \exp \left[i \frac{q}{\hbar} \int_{ s \to {\rm Pinhole 1 \to screen}} \! \! \! \! \! \! \! \! \! \! \! \! \! \! \! \! \! \! \! \! \! \! \! \! \! \! \! \! \! \! \! \! \! \! \! \! \! \! \left( \phi dt- {\bf A} \cdot d {\bf x} \right) \right] \nonumber \\
\psi_{2}' &=& \psi_{2} \cdot  \exp \left[i \frac{q}{\hbar} \int_{ s \to {\rm Pinhole 2 \to screen}} \! \! \! \! \! \! \! \! \! \! \! \! \! \! \! \! \! \! \! \! \! \! \! \! \! \! \! \! \! \! \! \! \! \! \! \! \! \!  \left( \phi dt- {\bf A} \cdot d {\bf x} \right) \right] 
\label{eq:Interfere1}
\end{eqnarray}

where, $ \psi_{1}' $ and $ \psi_{2}' $ are the electron wave functions on the screen passing through pinhole 1 and 2 with the unobservable potentials respectively. $ \psi_{1} $ and $ \psi_{2} $ are the electron wave functions heading to pinhole 1 and 2 at the electron source without the effects of the unobservable potentials. $ \phi $ and $ {\bf A} $ include not only the unobservable potentials expressed as (\ref{eq:Maxi}) but also the unobservable part of the potentials generated by localized potentials such as (\ref{eq:AandC}) and (\ref{eq:Vectorpotential}). 

Then the probability of finding the electron on the screen by using these wave functions can be described as follows,
\begin{eqnarray}
P_{12} & \propto & | \psi' |^{2} = | \psi' _{1} + \psi'_{2} |^{2} \nonumber \\
& = & | \psi_{1} |^{2} + | \psi_{2} |^{2} - 2 {\rm Re}  \left( \exp \left[ i \frac{q}{\hbar} \oint_{ s {\rm \to 1 \to screen \to 2 \to} s} \! \! \! \! \! \! \! \! \! \! \! \! \! \! \! \! \! \! \! \! \! \! \! \! \! \! \! \! \! \! \! \! \! \! \! \! \! \left( \phi dt- {\bf A} \cdot d {\bf x} \right) \right] \psi_{1}^{*} \psi_{2} \right)
\label{eq:interfere2}
\end{eqnarray}

where 1 and 2 of the integration path denote pinhole 1 and 2 respectively. This description is identical to Aharonov-Bohm effect \cite{ABeffe}.

In case of single electron interference, we can find the electron at pinhole 1 without fail but not at pinhole 2, i.e., $ | \psi_{1} |^{2} = 1 $ and $ | \psi_{2} |^{2} = 0 $. Although the exact expression should be $ \int | \psi_{1 {\rm or} 2} |^{2} d {\bf V} = $ 1 or 0 instead of the probability densities, we continue analysis with $ | \psi_{1} |^{2} = 1 $ and $ | \psi_{2} |^{2} = 0 $ for simplicity.

When we introduce a phase difference $ \theta $ between $ \psi_{1} $ and $ \psi_{2} $, $ P_{12} $ expresses the interference as follows,
\begin{equation}
P_{12} \propto 1 - 2 {\rm Re}  \left( \exp i \left[ \theta_{AB} + \theta \right] \psi_{1}^{*} \psi_{2} \right)
\label{eq:interfere3}
\end{equation}
where $ \displaystyle \theta_{AB} =  \frac{q}{\hbar} \oint_{ s {\rm \to 1 \to screen \to 2 \to} s}  \! \! \! \! \! \! \! \! \! \! \! \! \! \! \! \! \! \! \! \! \! \! \! \! \! \! \! \! \! \! \! \! \! \! \! \! \! \left( \phi dt- {\bf A} \cdot d {\bf x} \right) $.
 
Note that when $ \theta $ is fixed, the interference can be observed on the screen as a function of $ \theta_{AB} $, i.e., position on the screen. When $ \theta_{AB} $ is fixed, the interference can be observed on a fixed position of the screen as a function of $ \theta $.

However, the wave function $ \psi_{2} $ as a probability density must satisfy incoherent expressions, i.e., $ \psi_{1}^{*} \psi_{2} \neq 0 $ and $ | \psi_{2} |^{2} = 0 $.

Then in order to clarify the exact representation, we introduce the electron number states $ | n \rangle $ that means there are $ n $ electrons and charge operator $ {\bf Q} \equiv \int d^{3} x j_{0} ( x ) $ defined by a conserved current $ j_{\mu} = ( q , {\bf i} ) $, i.e., $ \partial^{\mu} j_{\mu} = \frac{\partial q}{\partial t} + \nabla \cdot {\bf i} = 0 $. The charge operator satisfies $ {\bf Q} | n \rangle = n q  | n \rangle $, which means the {\it n} electron state is the eigenstate of $ {\bf Q} $ \cite{Karlson,QFT}.

Because the charge operator is defined by a conserved current which satisfies the Maxwell equations and $ {\bf Q} $ will corresponds to the expression of photon number operator $ {\bf n} = \hat{A}^{\dagger} \hat{A} $ from the viewpoint of derivation of the charge or photon numbers, we regard $ {\bf Q} $ as combinations of indefinite metric operators similar to (\ref{eq:a2}), i. e.,
\begin{eqnarray}
{\bf Q} & = & \hat{q}_{1}^{\dagger} \hat{q}_{1} \nonumber \\
\hat{q}_{2} & = & \frac{1}{2} \gamma e^{i \theta/2} \hat{q}_{1} - \frac{1}{2} \gamma e^{- i \theta/2} \hat{q}_{1} \nonumber \\
\hat{q}_{2}^{\dagger} & = & \frac{1}{2} \gamma e^{- i \theta/2} \hat{q}_{1}^{\dagger} - \frac{1}{2} \gamma e^{ i \theta/2} \hat{q}_{1}^{\dagger}
\label{eq:q2}
\end{eqnarray}
Then we can obtain the single electron interference as same manner as (\ref{eq:inter}) in Heisenberg picture, i. e.,
\begin{equation}
\langle I \rangle = \langle \psi | \left( \hat{q}_{1}^{\dagger} + \hat{q}_{2}^{\dagger} \right) \left( \hat{q}_{1} + \hat{q}_{2} \right) | \psi \rangle = q \left\{ \frac{1}{2} + \frac{1}{2} \cos \theta \right\}
\label{eq:Q-int-Heisen}
\end{equation}
where $ \langle I \rangle $ is the expectation value of charge intensity.

Similarly, the interference of Schr\"odinger picture can be calculated as follows.
\begin{eqnarray}
\langle I \rangle & =  & \left( \langle \psi_{1} | + \langle \psi_{2} | \right) {\bf Q}_{S}  \left( | \psi_{1} \rangle + | \psi_{2} \rangle \right)  \nonumber \\
& = & q + \langle \psi_{2} | {\bf Q}_{S} | \psi_{2} \rangle + q \langle \psi_{1}  |  \psi_{2} \rangle + q \langle \psi_{2} | \psi_{1} \rangle \nonumber \\
& = & q \left\{ \frac{1}{2} + \frac{1}{2} \cos \theta \right\} 
\label{eq:Q-int}
\end{eqnarray}
where the charge operator $ {\bf Q}_{S} $ and state $ | \psi_{1} \rangle $ of Schr\"odinger picture are obtained from $ {\bf Q} = \hat{q}_{1}^{\dagger} \hat{q}_{1} = e^{i \hat{\mathcal{H}} t / \hbar} {\bf Q}_{S}  e^{- i \hat{\mathcal{H}} t / \hbar} $ and $  e^{- i \hat{\mathcal{H}} t / \hbar} | \psi \rangle = | \psi \rangle_{S} \equiv | \psi_{1} \rangle $ respectively.
Because $  {\bf Q}_{S} = e^{- i \hat{\mathcal{H}} t / \hbar} \hat{q}_{1}^{\dagger}  \hat{q}_{1} e^{i \hat{\mathcal{H}} t / \hbar} =  e^{- i \hat{\mathcal{H}} t / \hbar} {\bf Q} e^{i \hat{\mathcal{H}} t / \hbar} $, we define $ \hat{q}_{S} \equiv e^{-i \hat{\mathcal{H}} t / \hbar} \hat{q}_{1} e^{i \hat{\mathcal{H}} t / \hbar} $.
Then $ {\bf Q}_{S} = \hat{q}_{S}^{\dagger} \hat{q}_{S} $ and 
\begin{eqnarray}
\hat{q}_{2} | \psi \rangle & = & e^{i \hat{\mathcal{H}} t / \hbar} \hat{q}_{S} \left( \frac{1}{2} \gamma e^{i \theta/2} - \frac{1}{2} \gamma e^{- i \theta/2} \right) e^{- i \hat{\mathcal{H}} t / \hbar} | \psi \rangle \nonumber \\
& = & e^{i \hat{\mathcal{H}} t / \hbar} \hat{q}_{S} \left( \frac{1}{2} \gamma e^{i \theta/2} - \frac{1}{2} \gamma e^{- i \theta/2} \right) | \psi \rangle_{S} \nonumber \\
& \equiv & e^{i \hat{\mathcal{H}} t / \hbar} \hat{q}_{S} | \psi_{2} \rangle
\label{eq:phi-schr}
\end{eqnarray}
Therefore state of $ | \psi_{1} \rangle$ and $ | \psi_{2} \rangle $ can be recognized as follows.

"an electron passes through pinhole 1 with the unobservable potentials" as  $ | \psi_{1} \rangle $ with $P_{1} = \langle \psi_{1} | \psi_{1} \rangle = \int | \psi_{1} |^{2} d {\bf V} = 1 $

and

"no electron passes through pinhole 2 with the unobservable potentials" as $ | \psi_{2} \rangle $ with $ P_{2} = \langle \psi_{2} |  \psi_{2} \rangle = \int | \psi_{2} |^{2} d {\bf V} =0 $.

In the above treatment, the new charge operator (\ref{eq:q2}) similar to (\ref{eq:a2}) is introduced in order to emphasize the identity of the mathematical formula. However, if we take advantage of direct product of the electron state $ | \psi \rangle $ and the vacuum photon state $ | 0 \rangle + | \zeta \rangle $ in Schr\"odinger picture, then a straightforward approach can be made as follows.

Traditional direct product of the electron state $ | \psi \rangle $ and the vacuum photon state $ | 0 \rangle $  is expressed as $ | \psi \rangle | 0 \rangle \equiv | \psi , 0 \rangle \equiv | \psi \rangle_{S} \equiv | \psi_{1} \rangle $.

From the above discussion, the vacuum photon state should be replaced by $ | 0 \rangle + | \zeta \rangle $ in Schr\"odinger picture. Therefore the direct product becomes $ | \psi \rangle \left( | 0 \rangle  + | \zeta \rangle \right) = | \psi , 0 \rangle  + |\psi , \zeta \rangle \equiv | \psi \rangle_{S} + | \psi , \zeta \rangle $. Because $ |\psi , \zeta \rangle = | \psi_{2} \rangle $, then the direct product becomes $  | \psi \rangle \left( | 0 \rangle  + | \zeta \rangle \right) = | \psi_{1} \rangle + | \psi_{2} \rangle $. This formula is identical with (\ref{eq:Q-int}).

From (\ref{eq:zeta}), we can define $ | \psi_{2} \rangle \equiv | \psi \rangle | \zeta \rangle = \left( \frac{1}{2} \gamma e^{i \theta/2} - \frac{1}{2} \gamma e^{- i \theta/2} \right) | \psi \rangle_{S} $, then (\ref{eq:q2}) and (\ref{eq:Q-int-Heisen}) can be obtained as follows.
\begin{eqnarray}
\langle I \rangle & =  & \left( \langle \psi_{1} | + \langle \psi_{2} | \right) {\bf Q}_{S}  \left( | \psi_{1} \rangle + | \psi_{2} \rangle \right)  \nonumber \\
& = & \langle \psi_{1} | \left( 1+ \frac{1}{2} \gamma e^{-i \theta/2} - \frac{1}{2} \gamma e^{i \theta/2} \right) {\bf Q}_{S} \left( 1 + \frac{1}{2} \gamma e^{i \theta/2} - \frac{1}{2} \gamma e^{- i \theta/2} \right) | \psi_{1} \rangle \nonumber \\
& = &  \langle \psi_{1} | \left( \hat{q}_{1}^{\dagger} + \hat{q}_{2}^{\dagger} \right) \left( \hat{q}_{1} + \hat{q}_{2} \right) | \psi_{1} \rangle
\label{eq:Q-int-2}
\end{eqnarray}

When we introduce the phase terms of (\ref{eq:Interfere1}) and (\ref{eq:interfere2}) as $ \theta_{1} $, $ \theta_{2} $ and $ \theta_{AB} = \theta_{1} -\theta_{2} $, the interference (\ref{eq:Q-int}) is calculated to be as follows.

\begin{eqnarray}
\langle I \rangle & =  & \left( e^{- i \theta_{1} } \langle \psi_{1} | + e^{-i \theta_{2}} \langle \psi_{2} | \right) {\bf Q}_{S}  \left( e^{i \theta_{1} } | \psi_{1} \rangle + e^{i \theta_{2}} | \psi_{2} \rangle \right)  \nonumber \\
& = & q + \langle \psi_{2} | {\bf Q}_{S} | \psi_{2} \rangle + q e^{ - i \theta_{AB} } \langle \psi_{1}  |  \psi_{2} \rangle + q e^{ i \theta_{AB} }  \langle \psi_{2} | \psi_{1} \rangle \nonumber \\
& = & q \left\{ \frac{1}{2} + \frac{1}{2} \cos \theta \right\} + q e^{ - i \theta_{AB}} \langle \psi_{1}  |  \psi_{2} \rangle + q e^{ i \theta_{AB}}  \langle \psi_{2} | \psi_{1} \rangle \nonumber \\
\label{eq:Q-int-AB}
\end{eqnarray}

Hence, $ \theta_{AB} $ does not seem to be the origin of the single electron interference. Aharonov-Bohm effect will be observed when there are substantial electrons in both pinholes. The single electron interference will originate from the unobservable potentials in vacuum $ |\psi , \zeta \rangle \equiv | \psi_{2} \rangle $.

The above discussion suggests that the state "no electron passes through pinhole 2 with the unobservable potentials" generates the phase difference (in other words, unobservable oscillatory field as mentioned above.) for the interference without electron charges.

In the above expression for $ | \psi_{2} \rangle $, there is no mathematical solution in usual Hilbert space. Therefore the state of "no electron passes through pinhole 2 with the unobservable potentials" must also be regarded as a vector with zero probability amplitude in indefinite metric Hilbert space as can be seen from (\ref{eq:Q-int}), (\ref{eq:phi-schr}) and we can express the quantum state of the interference without quantum-superposition state.

In case of the provisional treatment as described above, we can apply following operators and state to an arbitrary geometry instead of (\ref{eq:q2}) and (\ref{eq:phi-schr}).

\begin{eqnarray}
{\bf Q} & = & \hat{q}_{1}^{\dagger} \hat{q}_{1} \nonumber \\
\hat{q}_{2} & = & \gamma \sum_{j=1}^{\infty} r_{j} e^{i \theta_{j}} \hat{q}_{1} - \gamma \sum_{j=1}^{\infty} r_{j} e^{- i \theta_{j}} \hat{q}_{1} \nonumber \\
\hat{q}_{2}^{\dagger} & = & \gamma \sum_{j=1}^{\infty} r_{j} e^{- i \theta_{j}} \hat{q}_{1}^{\dagger} - \gamma \sum_{j=1}^{\infty} r_{j} e^{i \theta_{j}} \hat{q}_{1}^{\dagger}
\label{eq:q2-arb}
\end{eqnarray}

\begin{eqnarray}
\hat{q}_{2} | \psi \rangle & = & e^{i \hat{\mathcal{H}} t / \hbar} \hat{q}_{S} \left( \gamma \sum_{j=1}^{\infty} r_{j} e^{i \theta_{j}} - \gamma \sum_{j=1}^{\infty} r_{j} e^{- i \theta_{j}} \right) e^{- i \hat{\mathcal{H}} t / \hbar} | \psi \rangle \nonumber \\
& = & e^{i \hat{\mathcal{H}} t / \hbar} \hat{q}_{S} \left( \gamma \sum_{j=1}^{\infty} r_{j} e^{i \theta_{j}} - \gamma \sum_{j=1}^{\infty} r_{j} e^{- i \theta_{j}} \right) | \psi \rangle_{S} \nonumber \\
& \equiv & e^{i \hat{\mathcal{H}} t / \hbar} \hat{q}_{S} | \psi_{2} \rangle
\label{eq:phi-schr-arb}
\end{eqnarray}

By using these operators or state, the same expectation value (\ref{eq:arb-number}) is obtained.

Note that the calculation using the superposition state of (\ref{eq:interfere2}) will be justified in case of mixed state whose probability is statistical sense.

\section{Discussion}\label{sec:DC}
\subsection{uncertainty principle and the reduction of the wave packet}

By the existence of the unobservable (scalar) potentials, Heisenberg's uncertainty principle can be explained independently of measurements. In addition, the paradox of the reduction of the wave packet typified by "Schr\"odinger's cat" and "Einstein, Podolsky and Rosen (EPR)"  \cite{Schr-cat,EPR} can be solved, because the origins of both are quantum-superposition state.

Former results clarify the description that the states of path 1 and 2 or pinhole 1 and 2 by Schr\"odinger picture are defined when the system is prepared expressed as a substantial single photon or electron and the unobservable (scalar) potentials respectively and each state does not split off such as quantum-superposition state, which means there is no reduction of the wave packet.

"When the system is prepared" corresponds to immediately after the branching point of the optical MZI or the pinholes. Which path or pinhole does the photon or electron select is unpredictable but after the selection, the state is fixed instead of quantum-superposition state. The concept of these states is identical with mixed states rather than
pure states formed by quantum-superposition, which insists the concept of quantum-superposition state is not required for the description of the phenomenon.

As for Heisenberg's uncertainty principle, we can clearly recognize it as trade-offs derived from Fourier transform non-related to measurement, which correspond to the canonical commutation relation.

\subsection{zero-point energy}
The electric field operators obtained from traditional quantization procedure for quantum optics with Coulomb gauge have relationships with harmonic oscillator as follows. (We consider only x-polarized photon for simplicity.) 
\begin{eqnarray}
\hat{A}_{1} & = & \frac{1}{ \sqrt{ 2 \hbar \omega}} \left( \omega \hat{q} + i \hat{p} \right) \nonumber \\
\hat{A}_{1}^{\dagger} & = & \frac{1}{ \sqrt{ 2 \hbar \omega}} \left( \omega \hat{q} - i \hat{p} \right)
\label{eq:trad-a}
\end{eqnarray}
where $ \hat{q} $ and $ \hat{p} $ are position and momentum operators obeying the commutation relation $ [ \hat{q} , \hat{p} ] = i \hbar $. Hamiltonian of harmonic oscillator is expressed as follows.
\begin{equation}
\hat{\mathcal{H}} = \frac{1}{2} \left( \hat{p}^{2} + \omega^{2} \hat{q}^{2} \right)
\end{equation}
Then following relations are obtained.
\begin{eqnarray}
\hat{A}_{1}^{\dagger} \hat{A}_{1} & = &  \frac{1}{ 2 \hbar \omega} \left( \hat{p}^{2} + \omega^{2} \hat{q}^{2} + i \omega \hat{q} \hat{p} - i \omega \hat{p} \hat{q} \right) \nonumber \\
& = & \frac{1}{ \hbar \omega} \left( \hat{\mathcal{H}} - \frac{1}{2} \hbar \omega \right) \nonumber \\
\hat{A}_{1} \hat{A}_{1}^{\dagger} & = &  \frac{1}{ \hbar \omega} \left( \hat{\mathcal{H}} + \frac{1}{2} \hbar \omega \right)
\label{eq:hamil}
\end{eqnarray}

From (\ref{eq:hamil}) and $ \langle 0 | \hat{A}_{1}^{\dagger} \hat{A}_{1} | 0 \rangle = 0 $, traditional zero-point energy has been recognized as $ \langle 0 | \hat{\mathcal{H}} | 0 \rangle = \frac{1}{2} \hbar \omega $, i. e.,
\begin{eqnarray}
\langle 0 | \hat{A}_{1}^{\dagger} \hat{A}_{1} | 0 \rangle & = & \frac{1}{ \hbar \omega} \langle 0 | \left( \hat{\mathcal{H}} - \frac{1}{2} \hbar \omega \right)  | 0 \rangle \nonumber \\
& = & \frac{1}{ \hbar \omega} \left( \langle 0 | \hat{\mathcal{H}} | 0 \rangle 
- \frac{1}{2} \hbar \omega \right) = 0
\end{eqnarray}
This traditional fixed zero-point energy originates from the definition of the electric field operators in (\ref{eq:trad-a}) without the unobservable (scalar) potentials. However we have obtained the idea that there are unobservable potentials in whole space. Then we should replace (\ref{eq:trad-a}) with followings by using the operators in (\ref{eq:a2}).
\begin{eqnarray}
\hat{A}_{0}' + \hat{A}_{1} & = & \frac{1}{ \sqrt{ 2 \hbar \omega}} \left( \omega \hat{q} + i \hat{p} \right) \nonumber \\
\hat{A}_{0}'^{\dagger} + \hat{A}_{1}^{\dagger} & = & \frac{1}{ \sqrt{ 2 \hbar \omega}} \left( \omega \hat{q} - i \hat{p} \right) 
\label{eq:trad-rep}
\end{eqnarray}

Therefore Hamiltonian will be the same expression of the interference as follows.
\begin{equation}
\hat{\mathcal{H}} = \hbar \omega \left( \hat{A}_{0}'^{\dagger} \hat{A}_{0}' + \hat{A}_{1}^{\dagger} \hat{A}_{1} + \hat{A}_{0}'^{\dagger} \hat{A}_{1} + \hat{A}_{1}^{\dagger} \hat{A}_{0}' \right) + \frac{1}{2} \hbar \omega
\label{eq:hamil-rep2}
\end{equation}

Hence the energy of single photon state also fluctuates.
\begin{equation}
\langle 1 | \hat{\mathcal{H}} | 1 \rangle = \frac{1}{2}  \hbar \omega \langle 1 |  \hat{A}_{1}^{\dagger} \hat{A}_{1} |1 \rangle + \frac{1}{2}  \hbar \omega \langle 1 |  \hat{A}_{1}^{\dagger} \hat{A}_{1} |1 \rangle \cos \theta + \frac{1}{2} \hbar \omega
\label{eq:1photoene}
\end{equation}
Because  a single photon can be observed when $ \theta = \pm N \pi , \ (N : {\rm even \ numbers}) $, then
\begin{eqnarray}
\langle 1 | \hat{\mathcal{H}} | 1 \rangle & = & \frac{1}{2}  \hbar \omega \langle 1 |  \hat{A}_{1}^{\dagger} \hat{A}_{1} |1 \rangle + \frac{1}{2}  \hbar \omega \langle 1 |  \hat{A}_{1}^{\dagger} \hat{A}_{1} |1 \rangle + \frac{1}{2} \hbar \omega \nonumber \\
& = & \langle 1 |  \hat{A}_{1}^{\dagger} \hat{A}_{1} |1 \rangle \hbar \omega + \frac{1}{2}  \hbar \omega =  \hbar \omega
\label{eq:1photoene-0deg}
\end{eqnarray}
Therefore $  \langle 1 |  \hat{A}_{1}^{\dagger} \hat{A}_{1} |1 \rangle = \frac{1}{2} $ which leads to the replacement by following expectation values of photon number.
\begin{equation}
\langle 0 | \hat{A}_{1}^{\dagger} \hat{A}_{1} | 0 \rangle = -\frac{1}{2}, \ \ \langle 1 | \hat{A}_{1}^{\dagger} \hat{A}_{1} | 1 \rangle = \frac{1}{2}, \ \ \langle 2 | \hat{A}_{1}^{\dagger} \hat{A}_{1} | 2 \rangle = \frac{3}{2}, \ \ \cdots
\end{equation}
Traditionally, $ \langle 0 | \hat{A}_{1}^{\dagger} \hat{A}_{1} | 0 \rangle $ has been considered to be 0. However we should recognize $ \langle 0 | \hat{A}_{1}^{\dagger} \hat{A}_{1} | 0 \rangle = - \frac{1}{2} $ which requires indefinite metric.

Then absolute value of the single photon interference moves depending on the selection of  $ \langle 0 | \hat{A}_{1}^{\dagger} \hat{A}_{1} | 0 \rangle $. However $ \langle \hat{I} \rangle \propto \frac{1}{2} \pm \frac{1}{2} \cos \theta $ is maintained.

By using the expectation value, zero-point energy is calculated to be
\begin{eqnarray}
\langle 0 | \hat{\mathcal{H}} | 0 \rangle & = & \frac{1}{2} \hbar \omega \langle 0 | \hat{A}_{1}^{\dagger} \hat{A}_{1} | 0 \rangle + \frac{1}{2} \hbar \omega \langle 0 | \hat{A}_{1}^{\dagger} \hat{A}_{1} | 0 \rangle \cos \theta + \frac{1}{2} \hbar \omega \nonumber \\
& = & \frac{1}{4} \hbar \omega - \frac{1}{4}  \hbar \omega \cos \theta
\label{eq:zero-ene}
\end{eqnarray}
The zero-point energy also fluctuates, which can also explain spontaneous symmetry breaking. Note that if $ \hat{A}_{0}' = \gamma \hat{A}_{1} $, 
\begin{equation}
\hat{A}_{0}'^{\dagger} \hat{A}_{0}' = - \hat{A}_{1}^{\dagger} \hat{A}_{1} = - \frac{1}{ \hbar \omega} \left( \hat{\mathcal{H}} - \frac{1}{2} \hbar \omega \right)
\label{eq:hamil-rep}
\end{equation}
Hence the isolate indefinite metric potentials may possess negative energies \cite{Dirac}. However $ \hat{A}_{0}' \neq \gamma \hat{A}_{1} $ as can be seen from (\ref{eq:a2}) and can not be isolated but combined instead such as (\ref{eq:trad-rep}), the negative energies can only appear through the interference with the localized potentials that express the substantial photon. Therefore the infinite zero-point energy due to the sum of infinite degree of freedom is eliminated by (\ref{eq:zero-ene}) with $ \theta = \pm N \pi , \ (N : {\rm even \ numbers}) $.

When we use the formula of the tensor product (\ref{eq:appinter}) for the expression of the interference (\ref{eq:hamil-rep2}) instead of the provisional treatment (\ref{eq:a2}), the phase $ \theta $ of the above discussion in this subsection is $ \pi $ shifted. In addition, (\ref{eq:divid-inf}) can be used for the fluctuation of the zero-point energy, i. e., $ 0 \leq \langle 0 | \hat{\mathcal{H}} | 0 \rangle \leq \frac{1}{2} \hbar \omega $.

The zero-point energy has been measured through Casimir effect \cite{Casimir1,Casimir2,Casimir3,Casimir4,Casimir5}.
The following circumstance can be identified as a typical setup for the measurement of Casimir effect. From the discussion of (\ref{eq:divid}), if a certain space that is not ''real vacuum'' but ''ideal vacuum'' is prepared and a certain geometry, e. g., two parallel plates, is placed in the space, then the zero-point energy of the space and geometry are calculated to be $  \frac{1}{2} \hbar \omega $ and $ 0 \leq \langle 0 | \hat{\mathcal{H}} | 0 \rangle \leq \frac{1}{2} \hbar \omega $ respectively. Because the energy of the geometry is not exceed that of the space, the geometry is subjected to a compressive stress from the space .

This kind of attractive force of the geometry derived from the energy difference is identical with the basic concept of Van der Waals force which will be the origin of Casimir effect \cite{VanderW}.

Therefore the above calculation will not be inconsistent with Casimir effect.

\subsection{spontaneous symmetry breaking}

Traditional treatment of the spontaneous symmetry breaking, which explores the possibility of $ {\bf Q} | 0 \rangle \neq 0 $ or generally "$ | 0 \rangle $ is not an eigenstate of $ {\bf Q} $", needs an intricate discussion using Goldstone boson or Higgs boson \cite{QFT,PhysRev.127.965}. Where $ | 0 \rangle $ is vacuum state. 

However, the unobservable potentials eternally populate the whole of space as mentioned above and there are no electron at pinhole 2. Therefore the state of pinhole 2, $ | \psi_{2} \rangle $, can be identified as vacuum instead of $ | 0 \rangle $.
From the relation $ \langle \psi_{2} |  \psi_{2} \rangle = 0 $ as described above, if $ | \psi_{2} \rangle $ is an eigenstate of $ {\bf Q} $, i.e., $ {\bf Q} | \psi_{2} \rangle = \alpha | \psi_{2} \rangle$, then $ \langle \psi_{2} | {\bf Q} | \psi_{2} \rangle = \alpha \langle \psi_{2} | \psi_{2} \rangle = 0 $, where $ \alpha $ is an eigenvalue. However from the discussion under (\ref{eq:Q-int}), $ \langle \psi_{2} | {\bf Q} | \psi_{2} \rangle $ fluctuates between $ -q $ and 0 depending on the phase difference. Hence the vacuum $ | \psi_{2} \rangle $ is not an eigenstate of $ {\bf Q} $, which  expresses the spontaneous symmetry breaking. In addition to this discussion, the above zero-point energy, i.e., vacuum is not an eigenstate of $ \hat{\mathcal{H}} $, also expresses the spontaneous symmetry breaking.

In other words, there is no fluctuation, i. e., there is symmetry, in the continuum space with completely random phases as mentioned in section \ref{sec:ISP}. This space can be identified as a real vacuum. 
However the fluctuation gains entrance into the real vacuum when a particular geometry is introduced, i. e., the symmetry is broken in the literature.

The above discussion that the real vacuum is filled with potentials whose state exists under original ground state is identical with the spontaneous symmetry breaking using the analogy of superconductivity when we replace $ {\bf Q} $ or $ \hat{\mathcal{H}} $ with energy level reported by Y. Nambu and G. Jona-Lasinio \cite{Nambu1,Nambu2}. When the phase difference is fixed, the one vacuum is selected and the selection breaks symmetry of vacuum.

In addition, the spontaneous symmetry breaking by the unobservable (scalar) potentials (gauge fields) leads to mass acquire of gauge fields (Higgs mechanism) \cite{Higgs}.

Therefore the above discussion will not be inconsistent with traditional treatment of spontaneous symmetry breaking and the mass acquire mechanism.

\subsection{general treatment of single particle interferences}
From (\ref{eq:new-int-sch}) and (\ref{eq:Q-int}), the single particle interferences can be expressed as following manner.
\begin{eqnarray}
\langle I \rangle & =  & \left( \langle \phi | + \langle \zeta | \right) {\bf F}  \left( | \phi \rangle + | \zeta \rangle \right) \nonumber \\
& = & f + \langle \zeta | {\bf F} | \zeta \rangle + f \langle \phi |  \zeta \rangle + f  \langle \zeta | \phi \rangle
\label{eq:general}
\end{eqnarray}
Then when $ \langle \zeta | {\bf F} | \zeta \rangle + f \langle \phi |  \zeta \rangle + f  \langle \zeta | \phi \rangle = - \frac{1}{2} f + \frac{1}{2} f \cos \theta $, single particle interferences of $ {\bf F} $ by 2-path geometry, i.e., $ \langle I \rangle = f \left\{ \frac{1}{2} + \frac{1}{2} \cos \theta \right \} $ can be generated. Where $ {\bf F} $ is an arbitrary observable operator  of the particle, $ | \phi \rangle $ is an eigenstate of $ {\bf F} $, $ f $ is the eigenvalue of $ {\bf F} $ under state $ | \phi \rangle $ and $ | \zeta \rangle $ is an indefinite metric vector expressing unobservable potentials. In case of Maxwell equations as described in this paper, $ | \zeta \rangle $ is identified as commutative gauge fields (Abelian gauge fields). When we study multicomponent state $ | \phi \rangle $, $ | \zeta \rangle $ will be identified as non-commutative gauge fields (non-Abelian gauge fields) \cite{Yang1,Yang2,Weinberg,Utiyama}. However the above general treatment can be applied in both cases.

When $ {\bf F} $ is a number operator $ {\bf n } $ of the particle and $ | \phi \rangle $ is single particle state $ | \phi \rangle = | 1 \rangle $ in (\ref{eq:general}), the expectation value of the single particle number fluctuates, i.e.,
\begin{eqnarray}
\left( \langle 1 | + \langle \zeta | \right) {\bf n } \left( | 1 \rangle + | \zeta \rangle \right) & = & 1 + \langle \zeta | {\bf n} | \zeta \rangle + \langle 1 |  \zeta \rangle + \langle \zeta | 1 \rangle \nonumber \\
& = & \frac{1}{2} + \frac{1}{2} \cos \theta
\label{eq:general-exis}
\end{eqnarray}
In case of arbitrary geometry, the expectation value will be identical with (\ref{eq:arb-number}) in the same manner as (\ref{eq:a2-arb}), (\ref{eq:zeta-arb}), (\ref{eq:q2-arb}) and (\ref{eq:phi-schr-arb}) as follows.
\begin{equation}
\langle \hat{I} \rangle \propto - \left\{ ( r_{1}^{2} + r_{2}^{2} + \cdot + r_{M}^{2} ) + \sum_{j \neq k}^{M} r_{j} r_{k} e^{i ( \theta_{j} - \theta_{k})} \right\} + 1
\end{equation}

These kinds of self fluctuation of a particle will be consistent with neutrino oscillation \cite{PhysRevD.15.655,Feldman:2013vca}.

\section{Summary}\label{sec:CC}
There are some unresolved paradoxes in quantum theory.

If we take advantage of the tensor form or indefinite metric vectors as described in this paper, the paradoxes can be removed. In addition, it can explain the uncertainty principle independently of measurements, eliminate infinite zero-point energy and cause spontaneous symmetry breaking without complexity.

We should consistently introduce indefinite metric because Maxwell equations are wave equations in Minkowski space. When we introduce state vectors in Minkowski space, indefinite metric vectors are absolutely required. The required vector should be recognized not only as an auxiliary field but also as a real physical field just like a homodyne local oscillator or negative oscillatory field which is the root cause of single photon and electron interferences. 

The results insist the vacuum space is filled with the unobservable potentials which can eternally exist as waves and correspond to scalar potentials.
This mechanism can be spontaneously obtained by using tensor form.

This idea provides exactly the same calculation and experimental results by using quantum-superposition state because the scalar potential forms the oscillatory field and the substantial photon or electron moves in the field with the interferences as if the quantum-superposition state exists. In addition, the concept is based on an analogy from the expression of substantial localize electromagnetic fields or an electron and the unobservable scalar potentials instead of enigmatic quantum-superposition state that forces us to imagine a photon or an electron passes through the both paths or pinholes despite a photon or an electron can not be split off.

Hence even if the calculation and experimental results are exactly the same, the basic concept is completely different and such the enigmatic concept can not exist in nature which means the engineering application based on quantum-superposition will not be accomplished forever. (In addition, the engineering application based on entanglement will not be accomplished forever as discussed in other letter \cite{Morimoto-Entangle}.) 

Here again, even if the traditional quantum theory can give the same result as this reformulation, this doesn't prove that we still need the concept of quantum superposition or pure states "to describe nature", which are the fundamentals of enigmatic and paradoxical quantum science since the beginning of last century.
What the "quantum superposition" does is a mere nice exercise which contributes not at all to physics but misleads physical science.

It is like a matter of one-to-one correspondence in an inverse problem. From the shape and conditions of musical instruments (reality) we can estimate the sound (outcome) of the musical instruments. Vice versa, from the sound (outcome) we can estimate a number of sound sources such as the real musical instruments (reality) and fake musical instruments something like an electrical audio equipment (fake reality).
Even if the measurement results (sound or outcome) are the same, the real source (reality) is unique. It is obvious that the reformulation is reality, which is intuitive and rigorously derived from the reality, i. e., Maxwell equation, relativity and covariant field quantization. Then the traditional quantum theory is fake reality to explain the outcome with illusions such as "probability amplitudes" instead reality, which can not exclude a lot of enigmatic, paradoxical and "counter-intuitive" concept and interpretation in addition to inconsistency in relativity.

Furthermore, this reformulation will not be inconsistent with traditional treatment of Casimir effect, spontaneous symmetry breaking, the mass acquire mechanism and can be applied to non-Abelian gauge fields.

The superposition states are justified in case of mixed states whose probabilities are statistical sense. However, quantum-superposition state is not necessarily required in case of pure state whose probability is fundamental sense, though the superposition principle may be used as a nice mathematical tool to simplify analyses (mere nice exercise). Therefore, the traditional quantum theory can be regarded as a kind of statistical physics with ''probabilistic interpretation''. In contrast, the reformulated quantum theory in this paper can be regarded as a kind of deterministic physics without ''probabilistic interpretation''.

The incompleteness of "Quantum theory", which has been alerted by A. Einstein, will originates from lack of introduction of indefinite metric. 
Then the traditional quantum theory must be established on erroneous concept from the very beginning as A. Einstein continued claiming for his life \cite{EPR}.

The reformulated Quantum theory with introduction of indefinite metric as a physical substance will be complete. Quantum theory should be re-formulated by using  the indefinite metric as physical substance without ''probabilistic interpretation''.

M. Arndt and K. Hornberger have reviewed some testing of quantum mechanical superpositions \cite{Arndt}, we hope the results will be tested by those technologies. 







\section*{Acknowledgment}
The author would like to thank K. Sato, Dr. S. Takasaka and Dr. S. Matsushita for their helpful discussions.

%

\end{document}